\documentclass[12pt]{article}
\usepackage{amssymb,graphicx,color,epsfig}
\begin{document}

\baselineskip=.22in
\renewcommand{\baselinestretch}{1.2}
\renewcommand{\theequation}{\thesection.\arabic{equation}}
\newcommand{\klmt}{\mbox{K\hspace{-7.6pt}KLM\hspace{-9.35pt}MT}\ }

\begin{flushright}
{\tt arXiv:0707.0357}
\end{flushright}

\vspace{5mm}

\begin{center}
{{{\Large \bf Relativistic Dynamics\\ of\\[2mm] Multi-BPS D-vortices
and Straight BPS D-strings}}\\[14mm]
{Inyong Cho,~~Taekyung Kim,~~Yoonbai Kim,~~Kyungha Ryu}\\[2.5mm]
{\it Department of Physics and BK21 Physics Research Division, \\
Sungkyunkwan University, Suwon 440-746, Korea}\\
{\tt iycho,~pojawd,~yoonbai,~eigen96@skku.edu}}\\[5mm]
\end{center}

\vspace{10mm}

\begin{abstract}
Moduli space dynamics of multi-D-vortices from D2${\bar {\rm D}}$2
(equivalently, parallel straight D-strings from D3${\bar {\rm D}}$3)
is systematically studied. For the BPS D-vortices, we show through
exact calculations that the classical motion of randomly-distributed
$n$ D-vortices is governed by a relativistic Lagrangian of free
massive point-particles. When the head-on collision of two identical
BPS D-vortices of zero  radius is considered, it predicts either
90${}^{\circ}$ scattering or 0${}^{\circ}$ scattering equivalent to
180${}^{\circ}$ scattering. Since the former leads to a reconnection
of  two identical D-strings and the latter does to a case of their
passing through each other, two possibilities are consistent with
the prediction of  string theory. It is also shown that the force
between two non-BPS vortices is repulsive. Although the obtained
moduli space dynamics of multi-BPS-D-vortices is exact in classical
regime, the quantum effect of an F-string pair production should be
included in determining the probabilities of the reconnection and
the passing through for fast-moving cosmic superstrings.
\end{abstract}

\newpage

\setcounter{equation}{0}
\section{Introduction}

The development of D-branes and related string dynamics during the
last decade have affected much string cosmology. Recently D- and
DF-strings have attracted attentions~\cite{Copeland:2003bj} as new
candidates of cosmic superstrings~\cite{Witten:1985fp}. In
understanding cosmological implications of the D(F)-strings, the
description in terms of effective field theory (EFT) is
efficient~~\cite{VS,Kibble:2004hq}, which accommodates various
wisdoms collected from the Nielsen-Olesen vortices of Abelian Higgs
model~\cite{MS}. In case of the Nielsen-Olesen vortices or other
solitons, the derivation of the BPS limit for static
multi-solitons~\cite{Bogomolny:1975de} and their moduli space
dynamics~\cite{Shellard:1988zx} have been two important ingredients
in making the analysis tractable and systematic.

In this paper, we consider D- and DF-strings produced in the
coincidence limit of D3${\bar {\rm D}}$3 as codimension-two
nonperturbative open string degrees. In the context of type II
string theory, we have two reliable EFT's of a complex tachyon field
reflecting the instability of D${\bar {\rm D}}$ system. One is the
nonlocal action derived in boundary string field theory
(BSFT)~\cite{Kraus:2000nj}, and the other is Dirac-Born-Infeld (DBI)
type action~\cite{Sen:2003tm,Garousi:2004rd}. If we restrict our
interest to parallel straight D(F)-strings along one direction, the
one-dimensional stringy objects can be dimensionally reduced to
point-like vortices as the (cosmic) vortex-strings have been
obtained from the Nielsen-Olesen vortices in Abelian Higgs model.
Specifically, in the context of EFT, D0-branes from D2${\bar {\rm
D}}$2 have been obtained as D-vortex configurations in
(1+2)-dimensions~\cite{Jones:2002si,Kim:2005tw}. For such D0-branes,
their BPS limit has been confirmed by a systematic derivation of the
BPS sum rule and the reproduction of the descent relation for static
single D-vortex~\cite{Sen:2003tm} and
multi-D-vortices~\cite{Kim:2006xi,Go:2007fv}.

Various dynamical issues on D- and DF-strings have been addressed
extensively in various contexts, for example, the collisions of
DF-strings~\cite{Jackson:2004zg}, the reconnection and formation of
Y-junctions~\cite{Copeland:2003bj,Copeland:2006eh}, the evolutions
of cosmic DF-string network~\cite{Tye:2005fn}, and the production of
D(F)-strings~\cite{Sarangi:2002yt}. Since the BPS limit is now
attained for static multi-D-vortices from D2${\bar {\rm D}}$2 (or
parallel straight multi-D-strings from D3${\bar {\rm D}}$3) in the
absence of supersymmetry, the systematic study of related dynamical
questions becomes tractable. The first step is to construct the
classical moduli space dynamics for randomly-distributed $n$
D-vortices involving their scattering~\cite{MS}.

In this paper, starting from the field-theoretic static BPS and
non-BPS multi-D(F)-vortex configurations, we derive systematically
the moduli space dynamics for $2n$ vortex positions. The Lagrangian
for randomly-distributed moving BPS D-vortices results in a
relativistic Lagrangian of $n$ free point-particles of mass equal to
the D0-brane tension before and after collision. The head-on
collision of two identical D-vortices of zero radius predicts either
90${}^{\circ}$ scattering, or 0${}^{\circ}$ (equivalently
180${}^{\circ}$) scattering different from the case of BPS vortices
with finite core size in the Abelian Higgs model~\cite{MS}. The
90${}^{\circ}$ scattering leads to reconnection of two colliding
identical D-strings~\cite{Shellard:1987bv}, and the 0${}^{\circ}$
scattering suggests another possibility that two D-strings pass
through each other. Two possibilities are now understood, but it
cannot determine probabilities of the reconnection, $P$, and the
passing, $1-P$, since our analysis is classical. The quantum
correction, e.g., the production of the F-string pairs, should be
taken into account in order to determine the
probabilities~\cite{Jackson:2004zg}. While all the previous moduli
space dynamics assumed a slow-motion~\cite{MS}, our result covers
the whole relativistic regime. Once the probabilities are borrowed
from the calculations of string theory, the result seems promising
for cosmological applications of superstrings in the sense that the
relativistic classical dynamics of BPS D-strings can proceed without
the help of numerical analysis. A representative example is the
formation and evolution of a cosmic string
network~\cite{Kibble:1984hp}, which has a significant cosmological
implication.

The rest of the paper is organized as follows. In section 2, we
introduce a DBI type effective action for D${\bar {\rm D}}$ system,
and briefly recapitulate the derivation of BPS limit of static
multi-D-vortices. In section 3, we derive the Lagrangian for the
four coordinates of $n=2$ BPS D-vortices of zero radius without
assuming a slow motion, and show 0${}^{\circ}$ scattering in
addition to 90${}^{\circ}$ scattering for the head-on collision of
two identical D-vortices. In section 4, we address the force between
two non-BPS D-vortices. We conclude with a summary of the obtained
results in section 5.

\setcounter{equation}{0}
\section{D${\bar {\bf D}}$ System and BPS Limit of Multi-D-vortices}

The properties of D$(p-2)$ (or ${\bar {\rm D}}(p-2)$) produced from
the system of D$p{\bar {\rm D}}p$ in the coincidence limit is
described by an EFT of a complex tachyon field, $T=\tau
\exp(i\chi)$, and two Abelian gauge fields of U(1)$\times$U(1) gauge
symmetry, $A_{\mu}$ and $C_{\mu}$. A specific form of a DBI type
action is~\cite{Sen:2003tm,Garousi:2004rd}
\begin{eqnarray}\label{ac}
S=-{\cal T}_{p}\int d^{p+1}x\,
V(\tau)\left[\,\sqrt{-\det(X^{+}_{\mu\nu})}
+\sqrt{-\det(X^{-}_{\mu\nu})}\,\,\right],
\end{eqnarray}
where ${\cal T}_{p}$ is the tension of the D$p$-brane, and
\begin{equation}\label{Xpm}
X^{\pm}_{\mu\nu}=g_{\mu\nu}+F_{\mu\nu}\pm C_{\mu\nu} +({\overline
{D_{\mu}T}}D_{\nu}T +{\overline {D_{\nu}T}}D_{\mu}T)/2.
\end{equation}
We use $F_{\mu\nu}=\partial_{\mu}A_{\nu}-\partial_{\nu}A_{\mu}$,
$C_{\mu\nu}=\partial_{\mu}C_{\nu}-\partial_{\nu}C_{\mu}$, and
$D_{\mu}T=(\partial_{\mu} -2iC_{\mu})T$ in what follows.

In this section, we shall briefly recapitulate the derivation of BPS
limit of static multi-D-strings (or DF-strings) from D3${\bar {\rm
D}}$3~\cite{Kim:2006xi}, which provides basic formulae of moduli
space dynamics of the BPS objects in the subsequent section. Since
the BPS limit is satisfied for parallelly-stretched D(F)-strings, we
shall only consider the motion and the collision of D-strings
keeping their parallel shape. Then the dynamics of parallel
one-dimensional D-strings in three dimensions reduces to that of
point-like D-vortices in two dimensions.

Concerned with the above discussion, let us take into account the
static multi-D-vortices. We also restrict our concern to the
D-vortices,
\begin{equation}\label{st}
T=T(x^{i}), \quad (i=1,2),
\end{equation}
without electromagnetic field, $F_{\mu\nu}=0$, throughout this
paper. The effect of electromagnetic field related with DF-strings
will be briefly discussed in conclusions.

Plugging (\ref{st}) in the stress components of the energy-momentum
tensor leads to
\begin{equation}
T^{i}_{\;j}=-\frac{2{\cal T}_2
V}{\sqrt{1+S_{mm}-\frac{1}{2}A_{mn}^2}}
\left[\delta_{ij}-\left(S_{ij}-\delta_{ij}S_{kk}\right)+\left(A_{ik}A_{jk}
-\frac{\delta_{ij}}{2}A_{kl}^2\right) \right], \label{tij}
\end{equation}
where
\begin{equation}\label{sij}
S_{ij}\left(A_{ij}\right) =\frac{1}{2}\left(\partial_i{\overline{
T}}\partial_j T\pm \partial_j{\overline{ T}}\partial_i T\right).
\end{equation}
Reshuffling the terms, the pressure difference can be written as
\begin{equation}
T^{x}_{\; x}-T^{y}_{\; y} = \frac{{\cal T}_3
V}{\sqrt{1+S_{ii}-\frac{1}{2}A_{ij}^2}}
\left[(\overline{\partial_{x}T+ i\partial_{y}T}) (\partial_{x}T-
i\partial_{y}T) +(\overline{\partial_{x}T- i\partial_{y}T})
(\partial_{x}T+ i\partial_{y}T) \right], \label{xxyy}
\end{equation}
which vanishes when the first-order Cauchy-Riemann equation is
satisfied,
\begin{equation}\label{BPS}
(\partial_{x}\pm i\partial_{y})T=0,\qquad (\,
\partial_x \ln \tau = \pm \partial_y \chi \;{\rm and} \;
\partial_y \ln \tau = \mp \partial_x \chi \,).
\end{equation}
Applying (\ref{BPS}) to the off-diagonal stress component $T^{x}_{\;
y}$, we confirm that it vanishes
\begin{eqnarray}
T^{x}_{\; y}&=&\frac{{\cal T}_3
V}{2\sqrt{1+S_{ii}-\frac{1}{2}A_{ij}^2}}
\left[(\overline{\partial_{x}T\pm i\partial_{y}T}) (\partial_{x}T\mp
i\partial_{y}T) -(\overline{\partial_{x}T \mp i\partial_{y}T})
(\partial_{x}T\pm i\partial_{y}T) \right]\\
&\stackrel{(\ref{BPS})}{=}&0.
\end{eqnarray}

Suppose that $n$ static D-vortices are located randomly in the
$(x,y)$-plane. The ansatz of the tachyon field is
\begin{eqnarray}\label{ans}
\displaystyle{T = \tau(x,y) e^{i \sum_{p=1}^n\theta_p},}\qquad
\theta_p = \tan^{-1} \frac{y-y_p}{x-x_p},
\end{eqnarray}
where ${\bf x}_{p}= (x_{p},y_{p})$ $(p=1,2,...,n)$ denotes the
position of each D-vortex. Inserting the ansatz (\ref{ans}) into the
Cauchy-Riemann equation (\ref{BPS}), we obtain the profile of the
tachyon amplitude,
\begin{equation}\label{tau}
\tau(x,y)=\prod_{p=1}^{n}\tau_{{\rm BPS}}|{\bf x}-{\bf x}_p|.
\end{equation}
Plugging the ansatz (\ref{ans}) and solution (\ref{tau}) into the
pressure components, we obtain $-T^{x}_{\; x}=-T^{y}_{\; y}=2{\cal
T}_2 V$. Only when we take the zero-radius limit of D-vortices,
$\tau_{{\rm BPS}}\rightarrow \infty$, the pressure components vanish
everywhere except the points where D-vortices are located,
$-T^{x}_{\; x}|_{{\bf x}={\bf x}_{p}}=-T^{y}_{\; y}|_{{\bf x}={\bf
x}_{p}} =2{\cal T}_2$, and the Euler-Lagrange equation of the
tachyon field is satisfied.

In the thin BPS limit with a Gaussian-type tachyon potential,
\begin{equation}\label{bsft}
V(\tau)=\exp\left(-\frac{\tau^{2}}{\pi R^{2}}\right),
\end{equation}
the computation of Hamiltonian for $n$ randomly-located D-vortices
(\ref{ans})--(\ref{tau}) reproduces the BPS sum rule,
\begin{eqnarray}
{\cal T}_{0}|n|= \int d^2 x \;{\cal H}_{\rm BPS}&=&2{\cal T}_2 \int
d^2x \; \lim_{\tau_{{\rm BPS}}\rightarrow \infty} V\left(\tau
\right) (1+S_{xx})
\label{ham}\\
&=& 2{\cal T}_2 \int d^2x \; \lim_{\tau_{{\rm BPS}}\rightarrow
\infty} V\left(\tau \right) S_{xx}
\label{sxc}\\
&=&2\pi^{2}R^{2}{\cal T}_{2}|n|,
\label{dsr}
\end{eqnarray}
where ${\cal T}_{0}$ denotes the mass of unit D-vortex. The last
line (\ref{dsr}) means that the descent relation for codimension-two
BPS branes, ${\cal T}_{0}=2\pi^{2}R^{2}{\cal T}_{2}$, is correctly
obtained. Note that, for $n$ superimposed D-vortices with rotational
symmetry, the integration in (\ref{ham}) yields the correct descent
relation without taking the infinite $\tau_{{\rm BPS}}$ limit (or
the BPS limit).

We have shown that the static multi-D-vortices in the limit of zero
radius have the following properties. First, the pressures,
$T^{x}_{\; x}$ and $T^{y}_{\; y}$, vanish everywhere except the
positions of D-vortices, and the off-diagonal stress, $T^{x}_{\;
y}$, vanishes completely. Second, the nontrivial D-vortex
configuration given by the solution to the first-order
Cauchy-Riemann equation also satisfies the Euler-Lagrange equation.
Third, with a Gaussian-type tachyon potential, the integrated energy
of static $n$ D-vortices shows that the BPS sum rule and the descent
relation for codimension-two BPS branes are correctly reproduced.
Therefore, the fulfillment of these necessary requirements suggests
that a BPS limit of multi-D-vortices from D3${\bar {\rm D}}$3 is
achieved, and that the Cauchy-Riemann equation can be identified
with the first-order Bogomolnyi equation. Since supersymmetry does
not exist in the D3${\bar {\rm D}}$3 system, the derivation of BPS
bound is lacked differently from the usual BPS vortices in Abelian
Higgs model. In this sense, the BPS properties of these multi-
D-vortices (or parallel D(F)-strings) need further study.

\setcounter{equation}{0}
\section{Moduli Space Dynamics of Multi-D-vortices}

Suppose that $n$ BPS D-vortices located randomly in the
$(x,y)$-plane start to move. It is known that the classical dynamics
of BPS multi-solitons is described in the context of moduli space
dynamics~\cite{MS,Shellard:1988zx}. Since the BPS D-vortices are
point-like objects of zero radius, the description in the moduli
space seems more natural  than the BPS Nielsen-Olesen vortices. In
order to construct a formalism of moduli space dynamics, we should
first identify the complete list of zero modes. Although we do not
study the complete list of zero modes of point-like BPS D-vortices
systematically by examining small
fluctuations~~\cite{Weinberg:1979er}, their arbitrary positions
${\bf x}_{p}$ in the $(x,y)$-plane should at least be a part of
those. Different from the usual theory of a complex scalar field
with spontaneously-broken global or local U(1) symmetry with a
finite vacuum expectation value of the Higgs
field~\cite{Davis:1985pt}, this tachyon effective action (\ref{ac})
with a runaway tachyon potential (\ref{bsft}) has infinite vacuum
expectation value of the tachyon amplitude and then supports neither
a gapless Goldstone mode nor gauge bosons with finite mass. This
reflects nonexistence of perturbative open string degrees after the
D${\bar {\rm D}}$ system decays~\cite{Sen:2004nf}.

The objects of our consideration are BPS codimension-two D-vortices
(D0-branes) of which classical dynamics is depicted by the motion of
$n$ point particles in two-dimensions. The BPS nature predicts a
free motion when they are separated, so the interaction exists only
in the range of collisions, $\tau_{{\rm BPS}}|{\bf x}_{p}-{\bf
x}_{q}|\le 1$ for $p\ne q$. Let us consider the moduli space
dynamics in two classes. One is for the D-vortices of which
inter-distances are larger than the size of each D-vortex,
$\tau_{{\rm BPS}}|{\bf x}_{p}-{\bf x}_{q}|> 1$, and the other is for
colliding D-vortices in the range of $\tau_{{\rm BPS}}|{\bf
x}_{p}-{\bf x}_{q}|< 1$.

We consider moduli space dynamics assuming that the time-dependence
of fields appears in the D-vortex positions,
\begin{equation}\label{tde}
{\bf x}_{p}(t)=(x_{p}(t),y_{p}(t)).
\end{equation}
From the BPS property of point-like D-vortices, the tachyon amplitude
(\ref{tau}) dictates
\begin{eqnarray}
\tau =
\left\{%
\begin{array}{ll}
0, & \hbox{at each ${{\bf x}}_{p}$}\,, \\
\infty, & \hbox{elsewhere}. \\
\end{array}%
\right.
\end{eqnarray}

Since the BPS limit of D-vortices was attained in the absence of the gauge
field $C_{\mu}$ and $A_{\mu}$,
the Lagrangian of our interest from the action (\ref{ac}) is
\begin{eqnarray}
\lefteqn{L({\bf x}_{p}(t),{\dot {\bf x}}_{p}(t))}\nonumber\\
&=&\int d^{2}x\, {\cal L}(\tau,\chi,\partial_{\mu}\tau,\partial_{\mu}\chi)
\label{fla}\\
&=& -2{\cal T}_p \int d^{2}x V(\tau)\sqrt{(1+S_{xx})^{2}-
\left[(\partial_{t}\tau)^{2}+(\tau\partial_{t}\chi)^{2}
+\tau^{2}(\partial_{t}\tau\partial_{i}\chi-\partial_{i}\tau\partial_{t}\chi)^{2}
\right]} \, , \label{lag2}
\end{eqnarray}
where $S_{xx}$ is given in (\ref{sij}). Since we assumed that the
time-dependence appears only in the positions of D-vortices
(\ref{tde}), the time-derivatives of the tachyon amplitude
(\ref{tau}) and the phase (\ref{ans}) become
\begin{eqnarray}\label{tder}
\partial_{0}\ln\tau=
-\sum_{p=1}^{n}\frac{{\dot{\bf x}}_{p}(t)\cdot({\bf x}-{\bf
x}_{p}(t))}{\left|{\bf x}-{\bf x}_{p}(t)\right|^2},
\qquad
\partial_{0}\chi=
\sum_{p=1}^{n}\frac{\epsilon_{ij}{\dot{\bf x}_{p}^{i}}\,({\bf
x}-{\bf x}_{p}(t))^j}{\left|{\bf x}-{\bf x}_{p}(t)\right|^2}.
\end{eqnarray}
Plugging (\ref{tder}) with the solutions (\ref{ans})--(\ref{tau}) of
the first-order Bogomolnyi equation and with their spatial
derivatives into the Lagrangian (\ref{lag2}), we have
\begin{eqnarray}
L^{(n)}({\bf x}_{p}(t),{\dot {\bf x}}_{p}(t)) &=&-\frac{{\cal
T}_{0}}{\pi^{2}R^{2}} \int d^{2}(\tau_{{\rm BPS}}x)
\exp\left[-\frac{\left(\prod_{s_{1}=1}^{n}\tau_{{\rm BPS}} |{\bf
x}-{\bf x}_{s_{1}}|\right)^{2}}{\pi R^{2}}\right]
\nonumber\\
&&\hspace{-29mm}\times
\left[\frac{1}{\tau_{{\rm BPS}}^{2}}+\left(\prod_{s_{2}=1}^{n}\tau_{{\rm BPS}}
|{\bf x}-{\bf x}_{s_{2}}|\right)^{2}
\sum_{s_{3},s_{4}=1}^{n}\frac{\cos \theta_{s_{3}s_{4}}}{
\tau_{{\rm BPS}}|{\bf x}-{\bf x}_{s_{3}}|
\tau_{{\rm BPS}}|{\bf x}-{\bf x}_{s_{4}}|
}\right]
\nonumber\\
&&\hspace{-29mm}\times \sqrt{1-\frac{ \displaystyle{
\left(\prod_{s_{5}=1}^{n}\tau_{{\rm BPS}} |{\bf x}-{\bf x}_{s_{5}}|
\right)^{2}\sum_{p,q=1}^{n}\frac{\cos \theta_{pq}}{ \tau_{{\rm
BPS}}|{\bf x}-{\bf x}_{p}| \tau_{{\rm BPS}}|{\bf x}-{\bf x}_{q}|}}
}{ \displaystyle{ \frac{1}{\tau_{{\rm
BPS}}^{2}}+\left(\prod_{s_{6}=1}^{n}\tau_{{\rm BPS}} |{\bf x}-{\bf
x}_{s_{6}}| \right)^{2}\sum_{s_{7},s_{8}=1}^{n}\frac{\cos
\theta_{s_{7}s_{8}}}{ \tau_{{\rm BPS}}|{\bf x}-{\bf x}_{s_{7}}|
\tau_{{\rm BPS}}|{\bf x}-{\bf x}_{s_{8}}|} } }{\,\dot {\bf x}}_{p}
\cdot {\dot {\bf x}}_{q} }\; ,\label{tla}
\end{eqnarray}
where $\theta_{s_{i} s_{j}}$ is the angle between two vectors,
$({\bf x}-{\bf x}_{s_i})$ and $({\bf x}-{\bf x}_{s_j})$. For the
non-BPS D-vortices with finite $\tau_{{\rm BPS}}$, the integration
over ${\bf x}$ in (\ref{tla}) looks impossible to be performed in a
closed form except for the case of $n$ superimposed D-vortices,
${\bf x}_{1}={\bf x}_{2}= \cdots = {\bf x}_{n}$ and ${\dot {\bf
x}}_{1}={\dot {\bf x}}_{2}= \cdots ={\dot {\bf x}}_{n}$,
\begin{eqnarray}\label{1la}
L({\bf x}_{1},{\dot {\bf x}}_{1})=-n{\cal T}_{0}
\left(1+\frac{1}{\tau_{{\rm BPS}}^{2}}\right)
\sqrt{1-\frac{\tau_{{\rm BPS}}^{2}}{1+{\tau_{{\rm BPS}}^{2}}}
{\dot {\bf x}}_{1}^{2}}\,,
\end{eqnarray}
which is nothing but the Lagrangian of $n$ free relativistic
particles of mass ${\cal T}_{0}(1+\tau_{{\rm BPS}}^{2})/\tau_{{\rm
BPS}}^{2}$ moving with a velocity $\tau_{{\rm BPS}}\, {\dot {\bf
x}}_{1}/\sqrt{1+\tau_{{\rm BPS}}^{2}}\,$. If we take the BPS limit
of infinite $\tau_{{\rm BPS}}$, the mass and the velocity become
${\cal T}_{0}$ and ${\dot {\bf x}}_{1}$, respectively. The result in
this limit suggests a correct moduli space dynamics of
randomly-distributed BPS D-vortices.

The classical motion of separated BPS objects is characterized by no
interaction between any pairs of BPS solitons due to exact
cancelation. Since we did not assume a slow motion in deriving the
effective Lagrangian (\ref{tla}) from the field-theory one
(\ref{fla}), the first candidate for the BPS configuration is the
sum of $n$ relativistic free-particle Lagrangians with mass ${\cal
T}_{0}$. From now on we shall show that it is indeed the case. For
any pair of D-vortices, we may assume that the separation is larger
than the size of each D-vortex, which is of order of $1/\tau_{{\rm
BPS}}$. This assumption is valid everywhere for the BPS D-vortices
obtained in the zero-radius limit, $\tau_{{\rm BPS}}\rightarrow
\infty$, except for the instance of collision, which is to be
considered later.

The first static part in the Lagrangian (\ref{lag2}) becomes a sum
of $n$ $\delta$-functions in the BPS limit as given in
(\ref{ham})--(\ref{dsr}), which is the condition for BPS sum rule.
Substituting it into the Lagrangian (\ref{tla}) and taking
$\tau_{{\rm BPS}}\rightarrow \infty$ limit in the square root, we
obtain
\begin{eqnarray}
L^{(n)}({\bf x}_{p},{\dot {\bf x}}_{p})
&=&-{\cal T}_{0}\int d^{2}{\bar x}\sum_{s_{1}=1}^{n}
\delta^{(2)}({\bar {\bf x}}-{\bar {\bf x}}_{s_{1}})
\\
&&\hspace{-3mm} \times \sqrt{1- \frac{ \displaystyle{
\sum_{p=1}^{n}\hspace{-1mm}\prod\limits_{\scriptstyle s_{2}=1
\atop \scriptstyle (s_{2}\ne p)}^{n} \hspace{-1mm} ({\bar {\bf
x}}-{\bar {\bf x}}_{s_{2}})^{2}\,{\dot {\bar {\bf x}}}_{p}^{2}
+\hspace{-1mm}\sum\limits_{\scriptstyle p,q=1 \atop \scriptstyle
(p\ne q)}^{n}\hspace{-1mm} |{\bar {\bf x}}-{\bar {\bf
x}}_{p}||{\bar {\bf x}}-{\bar {\bf x}}_{q}|
\hspace{-1mm}\prod\limits_{\scriptstyle s_{2}=1 \atop \scriptstyle
(s_{2}\ne p,q)}^{n}\hspace{-1mm} ({\bar {\bf x}}-{\bar {\bf
x}}_{s_{2}})^{2}\cos\theta_{pq}\, {\dot {\bar {\bf
x}}}_{p}\cdot{\dot {\bar {\bf x}}}_{q}} } { \displaystyle{
\sum_{s_{4}=1}^{n}\hspace{-1mm}\prod\limits_{\scriptstyle s_{3}=1
\atop \scriptstyle (s_{3}\ne s_{4}) }^{n} \hspace{-1mm}({\bar {\bf
x}}-{\bar {\bf x}}_{s_{3}})^{2}
+\hspace{-1mm}\sum\limits_{\scriptstyle s_{4},s_{5}=1 \atop
\scriptstyle (s_{4}\ne s_{5})}^{n} \hspace{-1mm}|{\bar {\bf
x}}-{\bar {\bf x}}_{s_{4}}| |{\bar {\bf x}}-{\bar {\bf
x}}_{s_{5}}| \hspace{-2mm}\prod\limits_{\scriptstyle s_{3}=1 \atop
\scriptstyle (s_{3}\ne s_{4},s_{5})}^{n}\hspace{-2mm} ({\bar {\bf
x}}-{\bar {\bf x}}_{s_{3}})^{2}\cos\theta_{s_{4}s_{5}}} } }
\nonumber\\
&=&-{\cal T}_{0}\sum_{p=1}^{n}\sqrt{1-{\dot {\bf x}}_{p}^{2}}\; ,
\label{flg}
\end{eqnarray}
where ${\bar {\bf x}}\equiv \tau_{{\rm BPS}}{\bf x}$, ${\bar {\bf
x}_p}\equiv \tau_{{\rm BPS}}{\bf x}_p$, and ${\dot {\bar {\bf
x}}}_{p}\equiv d(\tau_{{\rm BPS}}{\bf x}_{p}) /d(\tau_{{\rm BPS}}t)
={\dot {\bf x}}_{p}$. The resulting Lagrangian (\ref{flg}) describes
$n$ relativistic free particles of mass ${\cal T}_{0}$ in the speed
limit $|{\dot {\bf x}}_{p}|\le 1$ as expected. It correctly reflects
the character of point-like classical BPS D-vortices of which actual
dynamics is governed by the relativistic field equation of a complex
tachyon $T(t,{\bf x})$. In addition, the size of each BPS D-vortex
approaches zero as $\tau_{{\rm BPS}}$ goes to infinity, and thus the
description in terms of the free Lagrangian (\ref{flg}) is valid for
any case of small separation between two D-vortices, i.e.,
$\lim_{\tau_{{\rm BPS}}\rightarrow \infty} \tau_{{\rm BPS}}|{\bf
x}_{p}-{\bf x}_{q}|\rightarrow \infty$ for $|{\bf x}_{p}-{\bf
x}_{q}|>0$ $(p\ne q)$.

Although the obtained result looks trivial, actually the
relativistic Lagrangian of multi-BPS objects (\ref{fla}) has never
been derived through systematic studies of moduli space dynamics.
Traditional methods of the moduli space dynamics of multi-BPS
vortices assume a slow motion of BPS solitons, and then read the
metric of moduli space~\cite{Shellard:1988zx,MS}. Therefore, its
relativistic regime is supplemented only by numerical analysis which
solves field equations directly.

As we mentioned earlier, the obtained relativistic Lagrangian
(\ref{fla}) of $n$ BPS D-vortices is free from perturbative open
string degrees due to the decay of unstable D${\bar {\rm D}}$. It
means that the classical dynamics of BPS D-vortices with nonzero
separation can be safely described by (\ref{fla}) and should be
consistent with the numerical analysis dealing with time-dependent
field equations. However, the full string dynamics dictates the
inclusion of F-string pairs between two D-strings and perturbative
closed string degrees from the decay of D${\bar {\rm D}}$, which may
affect the dynamical evolution of BPS D-vortices in quantum level.
One may also ask whether or not this derivation of the relativistic
Lagrangian of free particles is a consequence of DBI type action.
The specific question is how much the square-root form of DBI action
(\ref{ac}) affects the derivation. Although we do not have any other
example to compare, the Lagrangian (\ref{fla}) backs up the validity
of the DBI type action (\ref{ac}) as a tree-level Lagrangian.

Another characteristic BPS property appears in the scattering of BPS
objects. That is a head-on collision of two identical spinless BPS
vortices in Abelian Higgs model showing $90^{\circ}$
scattering~\cite{MS,Shellard:1988zx} which leads to the reconnection
of two identical vortex-strings~\cite{Shellard:1987bv}. On the other
hand, two identical  D-strings can also pass through each
other~\cite{Jackson:2004zg}, which distinguishes the cosmic
superstrings from the cosmic strings. From now on we study the
dynamics of multi-BPS D-vortices when they are overlapped at the
moment of collision, and address this intriguing question.

Let us discuss the head-on collision of two identical BPS D-vortices
in comparison to the Nielsen-Olesen vortices in their BPS limit. For
the Nielsen-Olesen vortices, the size is characterized by the
inverse of the Higgs scale $v$, and is finite $\sim 1/v$ in the BPS
limit ($\lambda =1$). On the other hand, the mass scale of the
Lagrangian (\ref{flg}) for $n$ D-strings per unit length is
characterized by the tension of the lower dimensional brane, ${\cal
T}_0$. Meanwhile the D-vortex size is determined by $1/\tau_{\rm
BPS}$ which becomes zero in the BPS limit. Therefore, the D-vortex
size is different from the theory scale in the BPS limit,
\begin{equation}
{1 \over \tau_{\rm BPS}}  \to 0 \ll {1\over
{\cal T}_0} .
\end{equation}
The scattering of zero-radius vortices exhibits  a very different
picture from that of finite-size vortices. In the scattering of
``classical" particles, the finite-size objects exhibit only one
possibility which is the bouncing-back head-on collision. However,
the zero-radius objects exhibit another possibility which is passing
through each other owing to the zero impact parameter.

When the ``quantum" concept of identical particles is taken into
account, the interpretation of the scattering picture becomes
somewhat different. For identical quantum particles, two particles
are indistinguishable in their coalescence limit. The particles are
simply superimposed, which is a solution of overlapped solitons
satisfying a nonlinear wave equation. As a result, a particle sees
only a half of the moduli space, so the moduli space for a particle
is not a complete ${\rm R}^2$ but a cone as shown in Fig. 1.

For the zero-radius vortices, the apex of the cone is sharp and thus
singular. At the moment of collision at the singular apex, the
scattering is unpredictable. What we can consider is only the
symmetry argument. There is a ${\rm Z}_2$ symmetry between the upper and
the lower quadrant of the moduli space which is required to be kept
before and after the collision. Considering the symmetry there are
only two possible scattering trajectories. A vortex which climbs up
the cone either overcomes the apex straightly, or bounces back. The
former corresponds to the 90${}^{\circ}$ scattering in the physical
space. Since the identical vortices are indistinguishable in the coalescence
limit (at the apex), it is unpredictable if the vortex has scattered
to the right or to the left. The latter bouncing-back case
corresponds to the 0${}^{\circ}$ (equivalently 180${}^{\circ}$) scattering
in the physical space. As Nielsen-Olesen vortices have a finite size
in the BPS limit, the moduli space is a stubbed cone of which apex
is smooth. The only possible geodesic motion of a vortex is
overcoming straightly the apex. Therefore, there is only the
90${}^{\circ}$ scattering in the physical space, and the
corresponding symmetry story is the same as for D-vortices.

The scattering story of vortices discussed so far can be continued
for two identical straight strings. The scattering of usual cosmic
strings mimics that of the Nielsen-Olesen BPS vortices. The
90${}^{\circ}$ scattering for vortices corresponds to the
``reconnection" for cosmic strings. Since this is the only
possibility for finite-size Nielsen-Olesen BPS vortices, the
reconnection probability is unity. Strings never pass through each
other.

The scattering picture of infinitely thin cosmic D-strings can be
borrowed from the scattering of the BPS D-vortices. In addition to
the reconnection as in cosmic strings, the cosmic D-strings can pass
through each other with a probability $1-P$, which corresponds to
the 0${}^{\circ}$ (180${}^{\circ}$) scattering for D-vortices.

The reconnection probability plays the key role in cosmologically
distinguishing cosmic strings and cosmic superstrings. Beginning
with the same initial configuration of the string network, cosmic
superstrings evolve in a different way from cosmic strings due to
non-unity $P$. Such a difference may be imprinted in the cosmic
microwave background and the gravitational wave radiation. In
addition, when F- and DF-strings are considered, a $Y$-junction can
be possibly formed.

The computation of the probability $P$ for cosmic superstrings
should be determined from string theory
calculations~\cite{Jackson:2004zg}. The F-string pair production
should also be included in determining $P$. When D-strings are
considered, there appears an F${\bar {\rm F}}$-string pair
connecting them. The energy cost of this pair production is
proportional to $2\ell$ where $\ell$ is the distance between
D-strings. In the coalescence limit $\ell \to 0$, the energy cost
becomes zero, so the F${\bar {\rm F}}$-pair arises possibly as
another zero mode of the theory. Note that this quantum level
discussion is beyond our classical analysis, but we can reproduce
the classical result: The scattering of two identical D-strings
stretched straightly to infinity, results in either reconnection or
passing through, which is different from the case of vortex-strings
based on the vortices of Abelian Higgs model.

\begin{figure}[t]
\begin{center}
\scalebox{0.8}[0.8]{\includegraphics{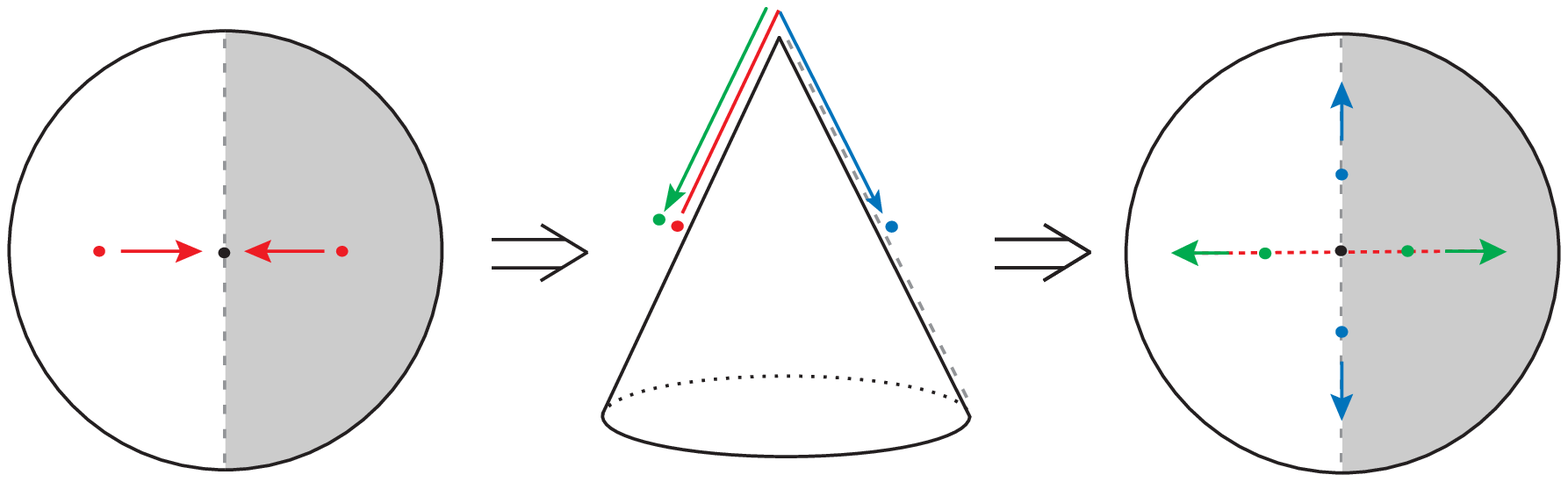}}
\scalebox{0.8}[0.8]{\includegraphics{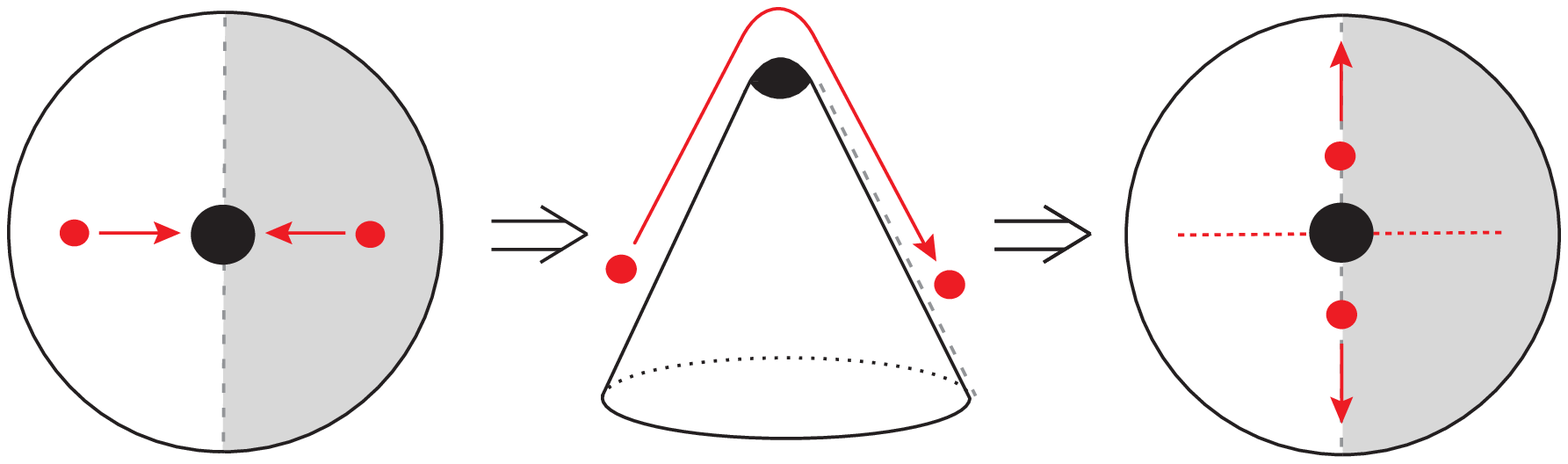}}
\par
\vskip-2.0cm{}
\end{center}
\caption{\small Scattering picture of two identical BPS D-vortices
(upper panel) and of BPS Nielsen-Olesen vortices (lower panel).
Since the two vortices are identical, a vortex sees only the half
space (the shaded region is equivalent to the unshaded region). In
addition, there is a ${\rm Z}_2$ symmetry between the upper and the
lower quadrant. After semi-diametric dashed lines are identified,
the moduli space for a vortex becomes a cone. For the zero-size
D-vortices, the moduli space is a sharp cone. Considering the
symmetry the scattering has two possibilities, 90${}^{\circ}$
scattering and 0${}^{\circ}$ (equivalently 180${}^{\circ}$)
scattering. For the finite-size Nielsen-Olesen vortices, the moduli
space is a stubbed cone, and there is only the 90${}^{\circ}$
scattering.}  \label{fig1}
\end{figure}

\setcounter{equation}{0}
\section{Interaction between non-BPS D-vortices}

In the previous section, we considered the motion and scattering of
BPS D-vortices in the context of moduli space dynamics. In the
present section, we consider non-BPS D-vortex configurations in
(\ref{ans})--(\ref{tau}) with a finite $\tau_{{\rm BPS}}$, and study
their dynamics and interaction between two D-vortices in the same
manner based on (\ref{tde}). Note that the non-BPS configurations
under consideration are given by the solutions
(\ref{ans})--(\ref{tau}) of the first-order Bogomolnyi equation
(\ref{BPS}). However, they are not exact solutions, but approximate
solutions of the Euler-Lagrange equation. Therefore, the validity of
forthcoming analysis is probably limited, and the obtained results
may only be accepted qualitatively.

Since the Lagrangian (\ref{tla}) was derived for the configurations
of arbitrary $\tau_{{\rm BPS}}$, it can also be employed in
describing the motion and the interaction between non-BPS
D-vortices. Here we restrict our interest to the case of two
D-vortices $(n=2)$ since it is sufficient without loss of
generality. As far as the dynamics of two D-vortices is concerned,
only the relative motion is physically meaningful. Adopting the
center-of-mass coordinates, we consider two identical non-BPS
D-vortices at initial positions, ${\bf d}_{0}/2=(d/2,0)$ and $-{\bf
d}_{0}/2=(-d/2,0)$. As time elapses, the motion of two D-vortices
with size $1/\tau_{{\rm BPS}}$ is described in terms of the
positions, ${\bf d}(t)/2$ and $-{\bf d}(t)/2$, with the linear
momentum being conserved. Introducing rescaled variables, ${\bar
{\bf d}}\equiv \tau_{{\rm BPS}} {\bf d}/(\sqrt{\pi}R)^{1/n}$ and
${\bar t}\equiv \tau_{{\rm BPS}} t/(\sqrt{\pi}R)^{1/n}$, the
complicated Lagrangian (\ref{tla}) reduces to a simple one of the
single particle
\begin{eqnarray}
L^{(2)}({\bar {\bf d}}(t),{\dot {\bar {\bf d}}}(t)) =-\frac{4{\cal
T}_{0}}{\pi}\int d^{2}{\bar x}\, e^{-
|{\bar {\bf x}}-{\bar {\bf d}}/2|^{2} |{\bar {\bf x}}+{\bar {\bf
d}}/2|^{2}} \left(d_{\tau}^{2}+{\bar {\bf x}}^2\right)
\sqrt{ 1-\frac{({\bar {\bf
d}}/2)^2}{d_{\tau}^{2}
+{\bar {\bf x}}^2} {\dot {\left(\frac{{\bar {\bf d}}}{2}\right)}}^{2} }\, ,
\label{pl2}
\end{eqnarray}
where $d_{\tau}\equiv 1/2\pi^{\frac{1}{4}}\sqrt{R}\,\tau_{{\rm BPS}}$
and ${\dot {\bar {\bf d}}}={\dot {\bf d}}$.

The first step to understand the mutual interaction between two
D-vortices is to investigate the potential energy,
\begin{eqnarray}
U^{(2)}({\bar {\bf d}}) &\equiv&-\left. L^{(2)}({\bar {\bf d}},{\dot
{\bar {\bf d}}})
\right|_{{\dot {\bar {\bf d}}}={\mathbf 0}}\\
&=&2{\cal
T}_{0}\left[1+\sqrt{\pi}\;d_{\tau}^{2}\;\exp\left(-\frac{{\bar {\bf
d}}^4}{32}\right) \emph{I}_{0}\left(\frac{{\bar {\bf
d}}^4}{32}\right)\right],\label{2po}
\end{eqnarray}
where $\emph{I}_{0}$ is the modified Bessel function.
\begin{figure}[h]
\begin{center}
\scalebox{1.3}[1.3]{\includegraphics{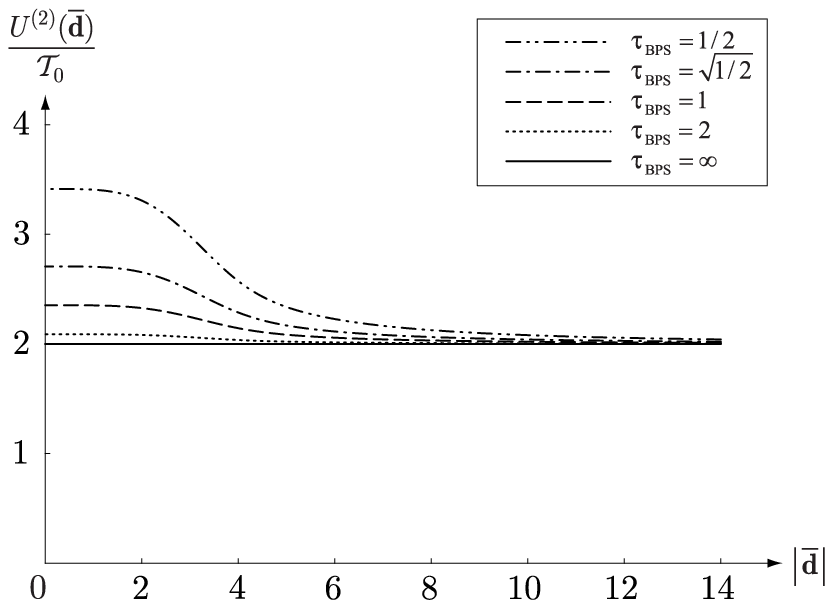}}
\par
\vskip-2.0cm{}
\end{center}
\caption{\small Potential energy $U^{(2)}({\tilde d})$ for various
$\tau_{{\rm BPS}}$'s.} \label{fig2}
\end{figure}
As shown in Fig.~\ref{fig2}, the first constant piece in (\ref{2po})
is independent of $\tau_{{\rm BPS}}$ and stands for the rest mass of
two D-vortices, $2{\cal T}_{0}$, in the BPS limit ($\tau_{{\rm
BPS}}=\infty$). The second distance-dependent term is a
monotonically-decreasing function. Its maximum value for the
superimposed D-vortices is $U^{(2)}({\bar {\bf d}}=0)-2{\cal T}_{0}=
{\cal T}_{0}/2R\tau_{{\rm BPS}}^{2}$ which vanishes in the BPS limit
of infinite $\tau_{{\rm BPS}}$. This $\tau_{{\rm BPS}}$-dependent
potential energy shows a repulsive short-distance interaction
between two non-BPS D-vortices. Fig.~\ref{fig2} also shows that as
$\tau_{{\rm BPS}}$ increases the system approaches the BPS limit
very rapidly.

In the current system, the conserved mechanical energy is nothing
but the Hamiltonian, $E^{(2)}({\bar {\bf d}},{\dot {\bar {\bf d}}})
=(\partial L^{(2)}/\partial {\dot {\bar {\bf d}}})\cdot {\dot {\bar
{\bf d}}}-L^{(2)}$. The kinetic energy is then given by
\begin{eqnarray}\label{2ki}
K^{(2)}({\bar {\bf d}},{\dot {\bar{\bf d}}}) &\equiv&
E^{(2)}({\bar {\bf d}},{\dot {\bar {\bf d}}})-
U^{(2)}({\bar {\bf d}})\nonumber\\
&=&\frac{4{\cal T}_{0}}{\pi}\int d^{2}{\bar x}\,
e^{-|{\bar {\bf x}}-{\bar {\bf d}}/2|^{2}
|{\bar {\bf x}}+{\bar {\bf d}}/2|^{2}}
\Bigg[\frac{1}{\sqrt{1-\frac{({\bar {\bf
d}/2)}^2}{d_{\tau}^{2}+{\bar {\bf x}}^2} {\dot
{\left(\frac{{\bar {\bf d}}}{2}\right)}}^{2}}}-1
\Bigg].
\end{eqnarray}
To understand the motion in detail, the spatial integration over
${\bar {\bf x}}$ should be performed for the kinetic term
(\ref{2ki}), but it is impossible when the function inside the
square root becomes negative,
\begin{equation}\label{ran}
0\le {\bar {\bf x}}^{2} < {\left(\frac{{\bar {\bf d}}}{2}\right)}^{2}
{\dot {\left(\frac{{\bar {\bf d}}}{2}\right)}}^{2} - d_{\tau}^{2}.
\end{equation}
If the distance $|{\bar {\bf d}}/2|$ and the speed $|{\dot {\bar
{\bf d}}}/2|$ are respectively smaller than the characteristic
length $d_{\tau}$ and the speed of light (unity in our unit system),
the integrand becomes imaginary and the moduli space dynamics is not
validly described anymore. As expected, for non-BPS D-vortices, this
formalism is applicable only to the regime of long distance and slow
motion, so-called the IR region. As $\tau_{{\rm BPS}}$ approaches
infinity in the BPS limit, $d_{\tau}$ becomes zero. Therefore, the
integration can be performed for all $|{\bar {\bf d}}/2|$ and
$|{\dot {\bar {\bf d}}}/2|$. (The UV physics is probed in the BPS
limit.)

It is necessary to consider the nonrelativistic limit of two
slowly-moving non-BPS D-vortices with $|{\dot {\bar {\bf d}}}/2|\ll
1$ in order to investigate the motions in detail. When the speed is
low enough, the nonrelativistic Lagrangian is given from (\ref{pl2})
as $L^{(2)}\approx (M^{(2)}/2)({\dot {\bar {\bf
d}}}/2)^{2}-U^{(2)}$. Here the reduced mass function $M^{(2)}$ is
\begin{eqnarray}\label{2ma}
M^{(2)}({\bar {\bf d}})=2{\cal T}_{0}\frac{\sqrt{\pi}}{4}\;{\bar {\bf
d}}^2\exp\left(-\frac{{\bar{\bf d}}^4}{32}\right)
\emph{I}_{0}\left(\frac{{\bar{\bf d}}^4}{32}\right).
\end{eqnarray}
\begin{figure}[t]
\begin{center}
\scalebox{1.3}[1.3]{\includegraphics{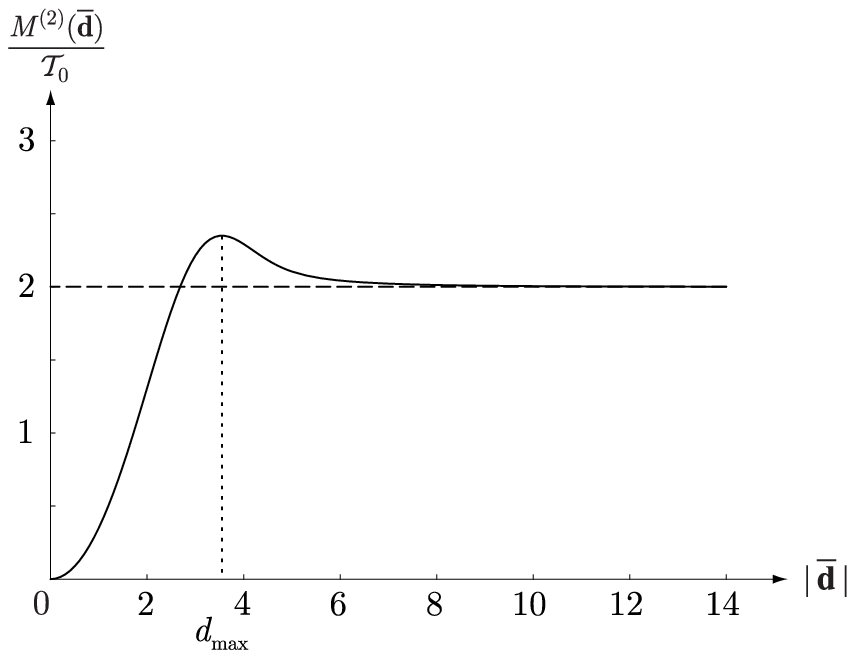}}
\par
\vskip-2.0cm{}
\end{center}
\caption{\small Mass function $M^{(2)}({\bar {\bf d}})$ of two
D-vortices in nonrelativistic motion.} \label{fig3}
\end{figure}
As shown in Fig.~\ref{fig3}, the mass function (\ref{2ma}) starts
from zero and increases to a maximum value at a finite $d_{{\rm
max}}$. Then it decreases rapidly and asymptotes to 2 at infinity.
Although the mass formula itself (\ref{2ma}) is independent of
$\tau_{{\rm BPS}}$, the inequality (\ref{ran}) puts a limit on the
validity. It is valid only at much larger distances than $d_{\tau}$.
Note also that the region of drastic mass change, where two
D-vortices are overlapped, should be excluded in reading detailed
physics.

The speed of a D-vortex is obtained from the nonrelativistic
mechanical energy $E^{(2)}$,
\begin{eqnarray}
\left|\;\frac{{\dot{\bar{\bf d}}}}{2}\;\right|=
\frac{1}{{\bar {\bf d}}}\sqrt{\frac{4(E^{(2)}/{\cal
T}_0-2)}{\sqrt{\pi}\exp\left(-\frac{{\bar{\bf d}}^4}{32}\right)
\emph{I}_{0}\left(\frac{{\bar{\bf
d}}^4}{32}\right)}-\frac{2}{\sqrt{\pi}R\tau_{{\rm BPS}}^2 }}\;.
\end{eqnarray}
When the initial speed $|{\dot{\bar{\bf d}}}_0/2|$ is lower than the
critical speed $1/\sqrt{2R\tau_{\rm BPS}^2}\;$ for non-BPS
D-vortices with finite $\tau_{\rm BPS}$, the D-vortex turns back at
a finite turning point due to the repulsive potential (\ref{2po}).
If the initial speed $|{\dot{\bar{\bf d}}}_0/2|$ exceeds the
critical speed, two identical D-vortices can collide at the origin
but this discussion in the region near the origin is not valid under
the nonrelativistic and long-distance approximation.

In studying the interaction and relative motion of identical non-BPS
vortices, we considered only two D-vortices. Extensions to the cases
of arbitrary number of non-BPS D-vortices are straightforward, at
least formally. Again, it should be noted that it is difficult to
perform explicitly the spatial integration for the
Lagrangian~(\ref{tla}).

\setcounter{equation}{0}
\section{Conclusions}

In this paper, we investigated the dynamics of D-strings produced in
the coincidence limit of D3${\bar {\rm D}}$3 as codimension-two
nonperturbative open string degrees. The model is described by a DBI
type effective action with a complex tachyon field. It was shown
in~\cite{Kim:2006xi} that the infinitely thin static tachyon profile
with a Gaussian-type potential reproduces the BPS configuration with
the correct BPS sum rule and descent relation. Since the D-strings
are parallelly stretched, their transverse dynamics is described by
point-like BPS D-vortices in two dimensions.

In this work, we investigated the dynamics of such $n$
randomly-distributed BPS D-vortices assuming that their positions
are time-dependent. We found that the classical moduli space
dynamics before and after collision is governed by a simple
Lagrangian $L^{(n)}({\bf x}_{p},{\dot {\bf x}}_{p}) = -{\cal
T}_{0}\sum_{p=1}^{n}\sqrt{1-{\dot {\bf x}}_{p}^{2}}$ which describes
$n$ free {\it relativistic} point particles with the mass given by
the D0-brane tension. Such a relativistic Lagrangian of multi-BPS
objects has never been derived through systematic studies of moduli
space dynamics. We also studied the classical scattering of
identical D-vortices. Different from the Abelian Higgs BPS vortices
with finite thickness, we could show that the head-on collision of
two identical D-vortices with zero radius exhibits either
90${}^{\circ}$ scattering or 0${}^{\circ}$ even in the relativistic
case.

Since the BPS limit is achieved in the zero-radius limit for
D-vortices, the obtained moduli dynamics possibly describes the
classical dynamics of the BPS D(F)-strings more accurately even for
the motion of high speed. Dynamics of cosmic superstrings can be
deduced analogously from the aforementioned vortex dynamics. After
the collision, the identical cosmic D-strings can either reconnect
with a probability $P$, or pass through without inter-commute with
$1-P$. The computation of the reconnection probability $P$ requires
string theory calculations. This picture is very different from that
of the usual Nielsen-Olesen cosmic strings which always reconnect
after the collision. In D-string collisions, the F-string pair
production should also be considered since the energy cost of such a
pair production becomes zero in the coincidence limit of D-strings.
This F-string pair may provide another zero mode in the scheme of
moduli space dynamics.

We studied the interaction of two D-vortices for the non-BPS case in
which the vortices have a finite size. We could show that the
effective potential for the motion exhibits a repulsive force.
Slowly incoming D-vortices will eventually bounce back. As the
vortex size approaches zero, the effective potential becomes
flatter, and eventually becomes completely flat which describes the
noninteracting BPS limit.

What we obtained shows a possibility for treating the BPS objects
and their dynamics from non-BPS systems without supersymmetry, and
thus further study to this direction is needed. In addition, it must
be an evidence for the validity of the DBI type effective action at
least in the classical level. Although the derivation of the
relativistic Lagrangian for free particles (\ref{fla}) seems
unlikely in the context of BSFT due to the complicated derivative
terms, it is worth tackling to check this point explicitly.

Our analysis is valid only for the straight D-strings and their
dynamics along transverse directions. Therefore, the next step is to
extend the analysis to the thin D(F)-strings of arbitrarily deformed
shapes. When an electric field ($E_z$) is turned on, D-strings
become DF-strings. When two DF-strings collide, they are known to
form a
Y-junction~\cite{Copeland:2003bj,Jackson:2004zg,Copeland:2006eh}.
Although the static BPS DF-configuration was obtained in the same
manner as the BPS D-strings~\cite{Kim:2006xi,Go:2007fv}, the
scattering of such DF-strings is probably more complicated, and will
not be simply described by what have been investigated here for
D-strings.

Although there do not exist perturbative zero modes from open string
side, massless closed string degrees are produced including
graviton, dilaton, and antisymmetric tensor field. These may affect
much on the dynamics of D(F)-strings as was done in the case of
global U(1) strings~\cite{Vilenkin:1981zs}. In a relation to the
string cosmology based on the \klmt setting, the BPS nature of
D(F)-strings in a warped geometry is an intriguing
subject~\cite{Firouzjahi:2006vp}.

\section*{Acknowledgments}
This work is the result of research activities (Astrophysical
Research Center for the Structure and Evolution of the Cosmos
(ARCSEC)) and grant No.\ R01-2006-000-10965-0 from the Basic
Research Program supported by KOSEF. This paper was supported by
Faculty Research Fund, Sungkyunkwan University, 2007 (Y.K.).

\end{document}